\documentclass[sigplan, 10pt]{acmart}
\setcopyright{rightsretained}
\copyrightyear{2023} 
\acmYear{2023} 

\acmConference[PPoPP '23]{The 28th ACM SIGPLAN Annual Symposium on Principles and Practice of Parallel Programming}{February 25-March 1, 2023}{Montreal, QC, Canada}
\acmBooktitle{The 28th ACM SIGPLAN Annual Symposium on Principles and Practice of Parallel Programming (PPoPP '23), February 25-March 1, 2023, Montreal, QC, Canada}
\acmDOI{10.1145/3572848.3577529}
\acmISBN{979-8-4007-0015-6/23/02}
\startPage{1}
\usepackage{booktabs}   
\usepackage{subcaption} 

\usepackage{amsmath,amsfonts}
\usepackage{algorithmic}
\usepackage{graphicx}
\usepackage{multirow}
\usepackage{stfloats}
\def\BibTeX{{\rm B\kern-.05em{\sc i\kern-.025em b}\kern-.08emT\kern-.1667em\lower.7ex\hbox{E}\kern-.125emX}}
\usepackage{makecell}
\usepackage{amsmath,amssymb,amsfonts}
\usepackage{algorithm}
\usepackage{algorithmic}
\usepackage{graphicx}
\usepackage{textcomp}
\usepackage{xcolor}
\def\BibTeX{{\rm B\kern-.05em{\sc i\kern-.025em b}\kern-.08em
    T\kern-.1667em\lower.7ex\hbox{E}\kern-.125emX}}
\newtheorem{definition}{Definition}
\newtheorem{theorem}{Theorem}
\newtheorem{lemma}{Lemma}
\begin{document}

\title{Lifetime-based Optimization for Simulating \\ Quantum Circuits on a New Sunway Supercomputer\\
}

\author{Yaojian Chen}
\affiliation{
\institution{Tsinghua University}
\country{Beijing, China}
}

\author{Yong Liu}
\affiliation{
\institution{National Supercomputing Center in Wuxi}
}
\affiliation{
\institution{Zhejiang Lab, Hangzhou, China}
}

\author{Xinmin Shi}
\affiliation{
\institution{Information Engineering University}
\country{Zhengzhou, China}
}

\author{Jiawei Song}
\affiliation{
\institution{National Supercomputing Center in Wuxi, China}
}

\author{Xin Liu}
\affiliation{
\institution{National Supercomputing Center in Wuxi}
}
\affiliation{
\institution{Zhejiang Lab, Hangzhou, China}
}

\author{Lin Gan}
\affiliation{
\institution{Tsinghua University}
}
\affiliation{
\institution{National Supercomputing Center in Wuxi, China}
}
\author{Chu Guo}
\affiliation{
\institution{Information Engineering University}
\country{Zhengzhou, China}
}
\author{Haohuan Fu}
\affiliation{
\institution{Tsinghua University}
}
\affiliation{
\institution{National Supercomputing Center in Wuxi, China}
}
\author{Jie Gao}
\affiliation{
\institution{National Research Centre of Parallel Engineering and Technology}
\country{Beijing, China}
}
\author{Dexun Chen}
\affiliation{
\institution{National Supercomputing Center in Wuxi, China}
}
\author{Guangwen Yang}
\affiliation{
\institution{Tsinghua University}
}
\affiliation{
\institution{National Supercomputing Center in Wuxi}
}
\affiliation{
\institution{Zhejiang Lab, Hangzhou, China}
}
\renewcommand{\shortauthors}{Chen, et al.}




\begin{abstract}
High-performance classical simulator for quantum circuits, in particular the tensor network contraction algorithm, has become an important tool for the validation of noisy quantum computing. In order to address the memory limitations, the slicing technique is used to reduce the tensor dimensions, but it could also lead to additional computation overhead that greatly slows down the overall performance. This paper proposes novel lifetime-based methods to reduce the slicing overhead and improve the computing efficiency, including an interpretation method to deal with slicing overhead, an in-place slicing strategy to find the smallest slicing set and an adaptive tensor network contraction path refiner customized for Sunway architecture. Experiments show that in most cases  the slicing overhead with our in-place slicing strategy would be less than the cotengra , which is the most used graph path optimization software at present. Finally, the resulting simulation time is reduced to 96.1s for the Sycamore quantum processor RQC, with a sustainable single-precision performance of 308.6Pflops using over 41M cores to generate 1M correlated samples, which is more than 5 times performance improvement compared to 60.4 Pflops in 2021 Gordon Bell Prize work.


\end{abstract}

\ccsdesc[500]{Software and its engineering~General programming languages}
\ccsdesc[300]{Social and professional topics~History of programming languages}

\keywords{quantum computing, circuit simulator tensor network contraction, Sunway architecture, slicing, direct memory access, fused design} 

\maketitle

\section{Introduction}

Quantum computer has the potential to provide exponential speedups over classical counterparts in specific tasks. The declaration ``Quantum Advantage" refers to those tasks that can only be solved in reasonable time by using quantum computers\cite{google-nature-2019}\cite{osti_1259664}\cite{zhong2020quantum}. However, despite such advantages in the computing capability, low fidelity is still the major challenge for quantum computers \cite{zlokapa2020boundaries}. Classical simulators are therefore important to provide validations for quantum computer design \cite{pan2021simulating}. Furthermore, scientists and researchers in areas that heavily rely on reliable computing resources, such as quantum algorithm, quantum programming language, and quantum compiler, can work on classical simulators and obtain close performance\cite{liu2021closing}. Considering both the exponential complexity and the urgent demand, improving the efficiency of the state-of-the-art classical simulator is important.

Directly storing an arbitrary quantum state on a classical computer requires an exponential amount of memory against the number of qubits. So using traditional state vector method to simulate quantum circuits is currently limited to less than 50 qubits\cite{de2007massively-11yearsbefore}. In recent years the tensor network contraction (TNC)\cite{google-unigraph} algorithm has demonstrated promising potential, especially in simulating random quantum circuits (RQCs). When combinated with the slicing technique, the TNC algorithm could efficiently simulate larger quantum circuits using only a limited amount of memory.

In TNC algorithm, qubits and quantum gates are represented as tensors, and the whole quantum circuit is treated as a tensor network \cite{biamonte2017tensor-nutshell}. The problem of computing amplitudes of the output quantum state is transformed into contracting the corresponding tensor networks. The performance of TNC is mainly determined by the TNC path. In a network with hundreds of tensors, a good contraction path can easily reduce the complexity by several magnitudes compared to the bad ones. In the meantime, during the contraction process, huge intermediate tensors could appear. The slicing technique can reduce the memory requirement\cite{villalonga2020establishing} \cite{breaking-49-qubit}, at the price of some computational overheads. This work aims to reduce the computational overhead caused by slicing.

Unlike previous efforts that predominantly use heuristic algorithms, this work introduces a new conception, \emph{lifetime}, in order to provide better interpretability for the slicing optimization in tensor network contraction. \emph{Lifetime} describes how an edge affects the TNC process by analyzing all the tensors and contractions it is involved in. By using \emph{lifetime}, one is able to transform the computation into an equivalent form with much less memory requirement. Slicing works on each two adjacent manually controllable levels on a multi-level storage system. For Sunway architecture (a most advanced supercomputing system in China), both the data exchange from hard disk to main memory, and from main memory to local data memory (LDM) (\emph{i.e.} process level and thread level), can be optimized by slicing. For process level, we slice tensors for distributed storage and parallelization, and apply \emph{lifetime}-based slice finder to reduce the overhead of process division. For thread level, slicing helps design a fused algorithm to reduce memory access, and transform the memory-intensive kernels into computationally intensive ones in some cases, thus improving the optimization capability.

Major contributions of this work include:
\begin{itemize}
\item A conception, \emph{lifetime} of graph edges (tensor dimensions), is proposed to explain the influence of each dimension, how slicing eliminates this influence, what is the origin of the slicing overhead, and how to avoid this overhead.
\item Guided by \emph{lifetime}, a dynamic slicing method with a new target function is proposed, to helps find smaller slicing set with less slicing overhead.
\item Indicated by \emph{lifetime}, a high performance algorithm for TNC that fits well with the Sunway architecture is presented.

\end{itemize} 

\section{Background}
\subsection{Tensor Network Contraction and Slicing}
\subsubsection{Notation}\label{notation}

A tensor network can be treated as an undirected graph, where tensors and dimensions are denoted by vertices and edges, respectively. We define the notation similar with the work of cotengra\cite{gray2021hyper}:

We denote a graph by $G = (V, E)$. $V$ is the vertex set, and $E$ is the edge set. The incident set $V_e$ of a edge $e$ contains the two vertices the edge connects. Map $w: E\rightarrow R$ denotes the edge weight, as well as the size on each dimension. Particularly in classical simulation, $w(e)$ is the power of $2$ for all $e \in E$. The incidence set $s_v$ of vertex $v$ is defined as $s_v = \{e: e \in E, v \in V_e \}$. Vertex contraction merges two vertex into one by removing shared edge and keeping the rest. In the generated graph $G^{'}$, the contracted vertices $v_0, v_1$ are replaced by a new vertex $v_2$, and the edges between $v_0$ and $v_1$ are removed in the new edge set. The uncontracted edges in $s_{v_0}$ and $s_{v_1}$ will be inherited by $s_{v_2}$.

\begin{figure}[htbp]
\centerline{\includegraphics[width=0.4\textwidth, height=0.3\textwidth]{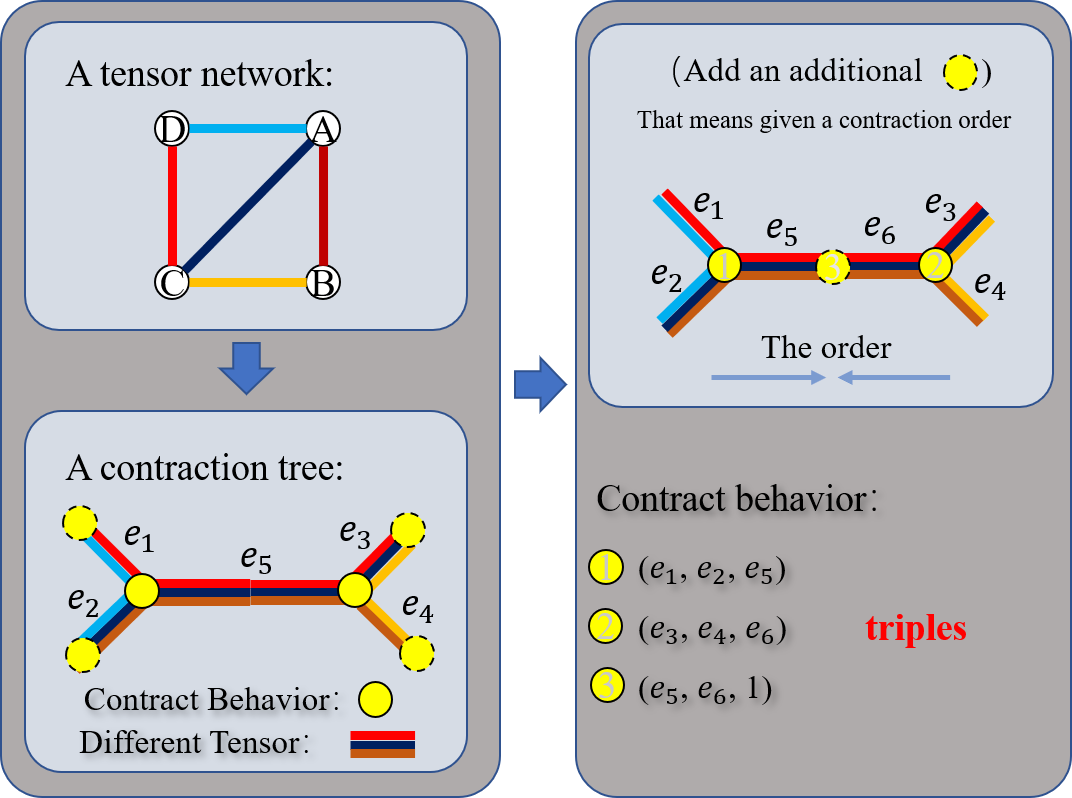}}
\caption{\textbf{Tensor network and its contraction tree. When a certain contraction path is formed, an additional contraction for the final scalar should be added.}}
\label{fig1}
\end{figure}

If we perform the pairwise contraction sequentially till there is only one vertex left, the order would form a contraction path. As a matter of fact, a equivalence class can be made up by reordering some independent steps, and all equivalent paths can be uniquely described by a tree structure. A contraction tree $B$ is defined as $B = (N_B, E_B)$, with each edge denoted by a vertex of original graph or an intermediate vertex generated by contraction. Specially, the edges connecting to leaf nodes are denoted by vertices in $V$. Nodes expressed as triplet $node = (e_1, e_2, e_3)$ refer to contractions. In particular, leaf nodes such as $(1, e_1, e_1)$ can be treated as a multiplication with a scalar $1$. Then leaf nodes can represent the reading process. A certain contraction path determines the direction of each contraction. By adding the final contraction as the root, as shown in Fig. \ref{fig1}, a rooted binary tree is acquired.

As for a given contraction tree, the time and space complexity of the corresponding contracting process can be quantitatively evaluated. The space cost is the biggest intermediate tensor, \emph{i.e} $\mathop{\max}_{v \in E_B} \prod_{e \in s_v}w(e)$. And the time complexity is:
\begin{equation}
C(B) = \sum_{(v_1, v_2, v_3) \in N_B} \prod_{e \in (s_{v_1} \cup s_{v_2} \cup s_{v_3})}w(e)
\end{equation}

To minimize the memory and time cost, the two expressions above are usually treated as the target function.

\begin{figure}[htbp]
\centerline{\includegraphics[width=0.46\textwidth, height=0.37\textwidth]{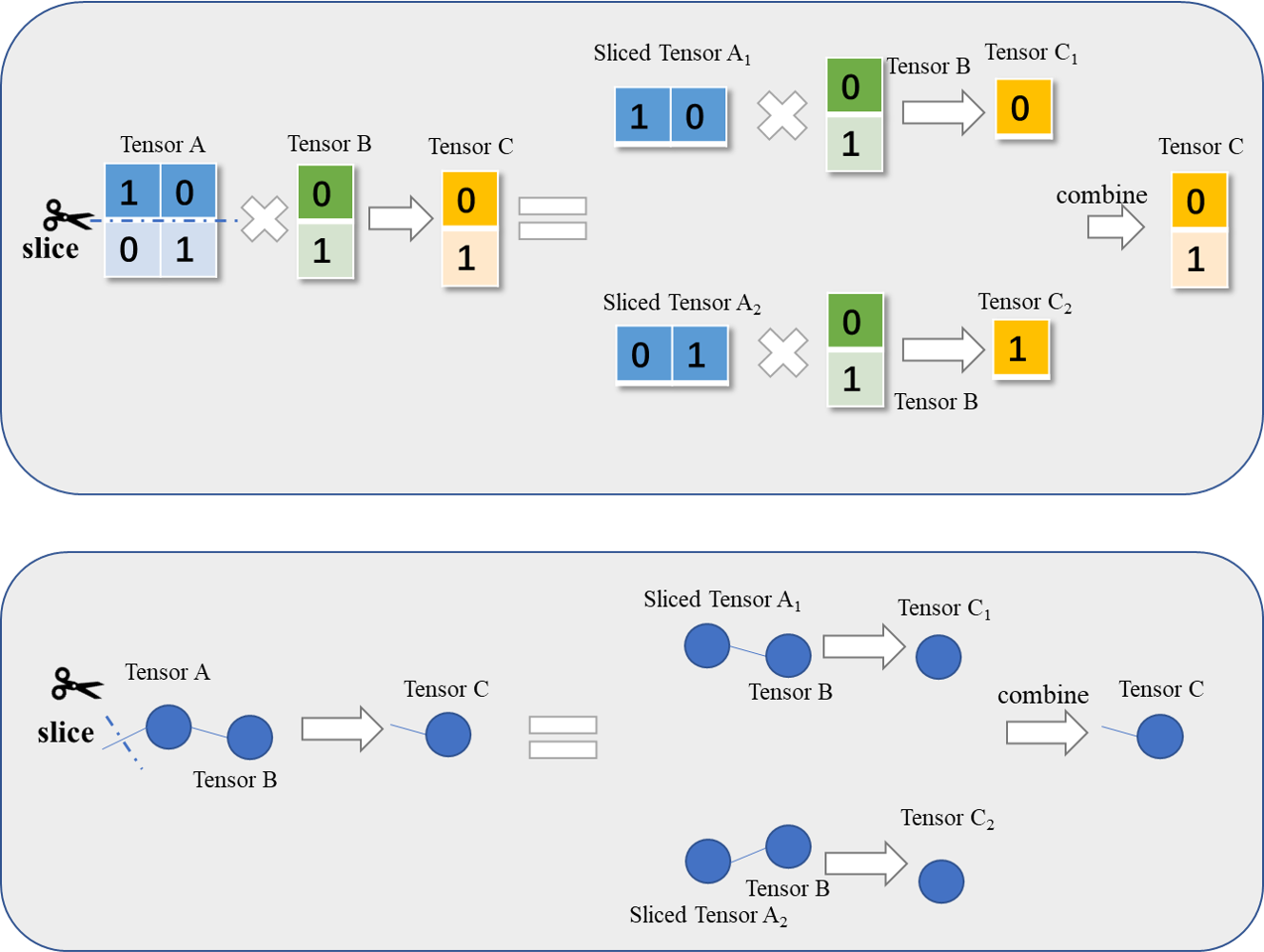}}
\caption{\textbf{An intuitive example of slicing. The second half expresses with graph as a parallel.}}
\label{fig2}
\end{figure}

For large tensor network, memory limitation is always a serious issue. So after a path is found, it is necessary to decide whether slicing is needed to reduce the dimension of tensors. Slicing a tensor alongside a dimension (an edge in the TN) can transform an $n$-dimension tensor into $w$ different $(n-1)$-dimension sub-tensors. Slicing temporarily cancels the correlation between these sub-tensors and provides parallelism. Fig.~\ref{fig2} shows a simple example of slicing. In this work, slicing are performed on the original $G$, in which there is $w(e) = 2$ for all $e \in E$. As a result, $s$ sliced edges make up $2^s$ independent tasks, and the result will be accumulated after their individual calculations. Total time complexity may increase after slicing under some circumstance, then slicing overhead is defined as:
\begin{equation}
O(B, S) = \frac{C_{slice}(B)\times2^{|S|}}{C_{original}(B)}
\end{equation}
where $S$ denotes the set of sliced edges. Slicing overhead comes from redundant calculations due to the split of tasks. From a higher perspective, slicing provides a distributed storage strategy and natural stream-like design since large tensors are sliced to stored and calculated in multi-processes, respectively. Compared to traditional methods based on heavy communication, slicing shows its advantage through independent subtasks. However, as the price, slicing overhead determines the effect, so one of the primary targets of this work is to explain and reduce such overhead.

\subsubsection{Related Work}
The demand for slicing originates from the considerable memory cost of high-rank tensors. In some quantum advantage circuits, there are tensors with dimensions of more than 60\cite{gray2021hyper}. They can occupy up to 1000 PB, which exceeds most storage systems. Slicing helps reduce the memory demand to the level of TBs or even GBs to make simulation feasible.

Cotengra\cite{gray2021hyper} integrates several anytime methods and claims over $10000\times$ speedup compared to the estimation made by Google\cite{google-nature-2019}. It introduced a series of  effective heuristic algorithms to search for contraction path, such as community\cite{girvan2002community} and graph partition\cite{akhremtsev2017engineering}. After the pre-process which is implemented in \cite{gray2018quimb}, rank-1 and rank-2 tensors are absorbed, and the tensor network is largely simplified. 
A greedy-based slicing strategy is built in contengra. It repeatedly chooses a dimension that leads to the most minor overhead to slice, until the memory demand is satisfied. Like most greedy methods, local minimum exists in this slicing strategy. In our work, contengra is applied to find a proper contraction path.

Alibaba proposed a simulator based on an observed structure called `stem'\cite{huang2020classical}. A stem is a computationally-intensive region in the contraction tree, including most of high-rank tensors. The inexpensive parts are called branches. Within a single stem, a bigger tensor sequentially absorbs smaller ones, and about $99\%$ computation cost happens during these contractions. For slicing, with a similar greedy strategy, they perform local tuning of contraction tree between two steps of slicing picking. This dynamic design highly reduced the inherent slicing overhead of a contraction tree. However, if the condition of local tuning is not satisfied, this strategy may not be able to find an optimal slicing set. This work can provide $10^{18.8}$ times complexity, slicing overhead of $4$, and a $14.7\%$ FLOPS efficiency. Due to the high efficiency, dynamic design is applied or updated by many other softwares, like cotengra\cite{gray2021hyper}.

As for the New Sunway System, several efforts study high-performance algorithms for TNC.
Sw\_Qsim\cite{li2021sw_qsim} provides a series of methods for matrix multiplication and tensor permutation, especially for narrow tensors. The work that won the 2021 ACM Gordon Bell prize\cite{liu2021closing} implemented Transpose Transpose GEMM Transpose (TTGT) algorithm on Sunway architecture, which is a fused design of tensor permutation and matrix multiplication in order to reduce memory access. These efforts significantly improve the FLOPS efficiency, but all concentrate on one step of the whole path instead of taking a holistic view. Optimization will soon reach the ceiling as a bandwidth-constrained problem according to Roofline model\cite{osti_1407078}. For extremely deep circuits, other tensor based methods like matrix product state are also implemented on Sunway architecture\cite{shangmps}. In this work, we focus on the hardware-motivated and highly entangled circuits, which have a clear 2D geometry and are relatively shallow. 

\subsection{Sunway 26010Pro Processor}

As a heavy time- and memory-consuming part, performing the actual contraction generally requires support from sophisticated supercomputers. A new Sunway supercomputer is selected for this work. 

\begin{figure}[htbp]
\centerline{\includegraphics[width=0.45\textwidth, height=0.3\textwidth]{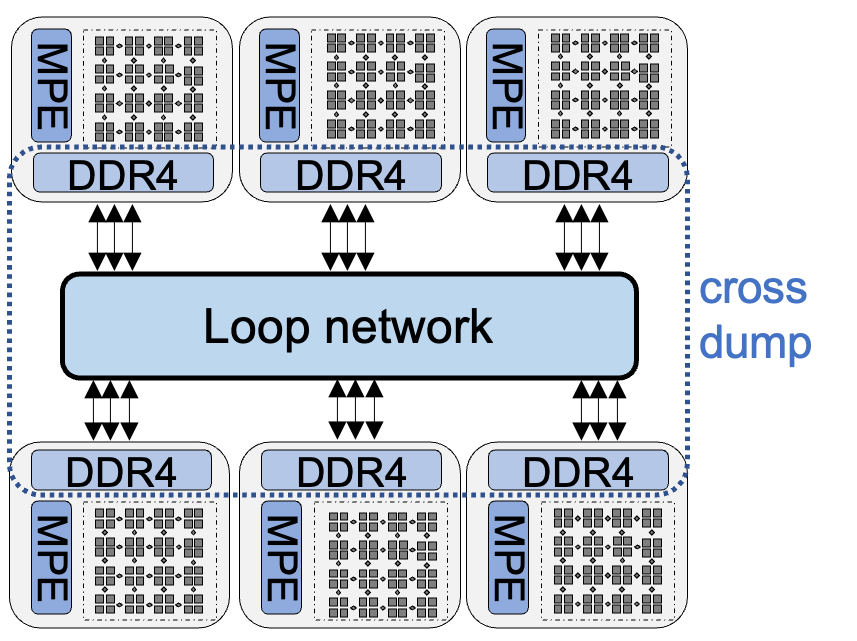}}
\caption{\textbf{SW26010pro Processor. The main memories of $6$ CGs are united to form a cross dump with $16\times 6$ GB in this work, in order to hold large tensors.}}
\label{fig3}
\end{figure}

The new-generation Sunway supercomputer has a similar architecture as Sunway TaihuLight\cite{fu2016sunway}, with the major computing capability provided by core groups (CGs). A heterogeneous many-core processor, SW26010pro, is designed for this supercomputer. Fig.~\ref{fig3} shows the structure of the chip. Each processor chip has 6 CGs. The computing processing elements (CPEs) of each CG are arranged as an 8 by 8 grid.

Each CG contains a 16GB main memory, and each CPE contains 256KB local data memory (LDM). Direct memory access (DMA) with a bandwidth of 51.2 GB/s is provided between LDM and main memory. Due to the enormous arithmetic intensity, the memory access bottleneck often turns to the vital problem for optimization. Remote memory access (RMA) with a peak bandwidth of more than 800 GB/s is designed for data exchange between CPEs within one CG.


\section{Lifetime\label{lifetime}}
\subsection{Definition}
\begin{figure}[htbp]
\centerline{\includegraphics[width=0.46\textwidth, height=0.18\textwidth]{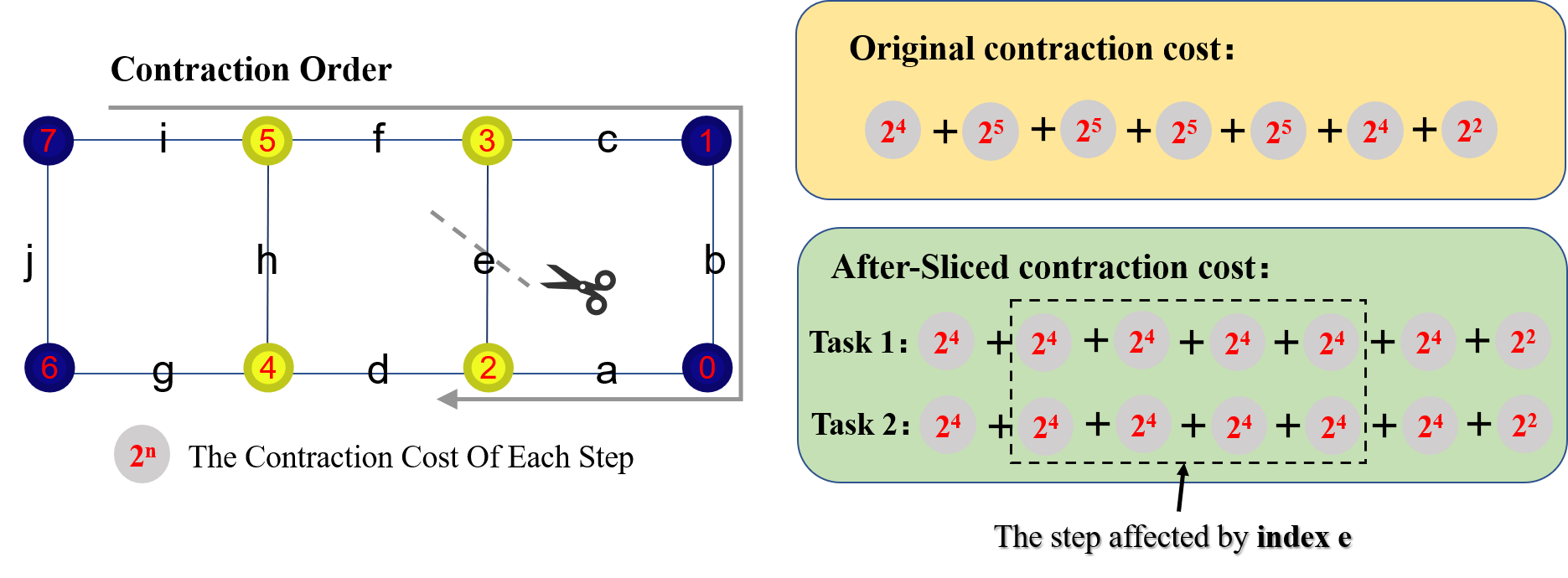}}
\caption{\textbf{An example to show the origin of slicing overhead. The left part is a tensor network represented by graph, and the arrow gives the contraction order. The right part shows the complexity before and after slicing. After slicing, the total complexity is the summation of all subtasks.}}
\label{fig4}
\end{figure}

Here is a simple example showing where the slice overhead comes from. Fig.~\ref{fig4} shows a contraction process on a $4 \times 2$ tensor network, and the target rank is limited to $3$. The ideal case is that the time complexity of all the contractions in one subtask becomes just $2^{-s}$ after slicing $s$ edges, and the original complexity is kept. However, from the right part, the first contraction and the last $2$ contractions in a subtask have not changed after sliced edge $e$, then they will be calculated twice. These redundant calculations make up the overhead. 

Actually, the sliced edge $e$ determines which contraction will be redundantly performed. Considering the contractions before slicing, edge $e$ involves four contractions:
\begin{equation}
\begin{aligned}
    \relax
    [f,h,j]\times[c,e,f]\rightarrow[c,e,h,j] \\
    [c,e,h,j]\times[b,c]\rightarrow[b,e,h,j] \\
    [b,e,h,j]\times[a,b]\rightarrow[a,e,h,j] \\
    [a,e,h,j]\times[a,d,e]\rightarrow[d,h,j]
\end{aligned}
\end{equation}
and these contractions are just inside the dotted box in ~Fig\ref{fig4}, which will not be redundantly computed.

Based on the analysis above, this work proposed a new concept, \emph{lifetime}, to describe the scope of influence of a sliced dimension during the slicing. The notation comes from section~\ref{notation}. 

\vspace{0.5em}
\begin{definition}
Given a tensor network $G = (V, E)$ and a contraction tree $B = (N_B, E_B)$, the concept \emph{lifetime} of a edge index $k \in E$ refers to a set of tensors $\{T_{i_1}, T_{i_2}, \dots, T_{i_n}\} \subset E_B$, if $k \in s_{T_{i_j}}$, for all $1\leq j\leq n$.
\end{definition}
\vspace{0.5em}

With the definition of \emph{lifetime}, the scope of influence of a sliced edge is clearly represented. We can conclude that, after slicing an edge $e$ from a tensor network, the size of tensors on the lifetime of $e$ will be halved while the size of the others will not change. The time complexity of the contractions corresponding to these tensors will not be changed, while the time complexity of the other contractions will be doubled.

Not merely on the whole contraction tree, \emph{lifetime} can be conveniently defined recursively on each sub-tree or intermediately generated tree during the contraction process by simply replacing the contraction tree in the definition. The advantage here is that we can focus more on the region with intensive computation. 

\subsection{Lifetime and Slicing Overhead}
\begin{figure}[htbp]
\centerline{\includegraphics[width=0.5\textwidth, height=0.39\textwidth]{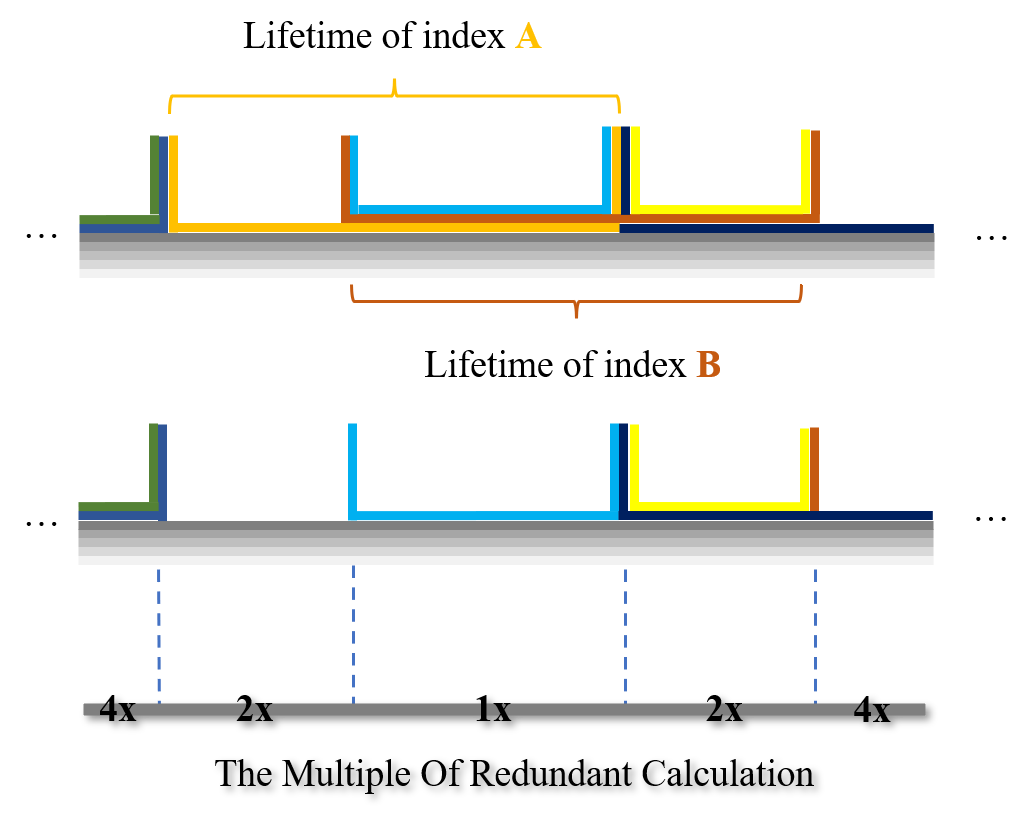}}
\caption{\textbf{The superposition rule of overhead. A contraction order is shown from left to right before (the upper half) and after (the lower half) slicing. Each branch denotes a tensor, and the contractions happen at the intersection points. Lines with different colors in the tensors denote the edges, and the collections of all the lines with the same color are naturally \emph{lifetime} of these edges.}}
\label{fig5}
\end{figure}

Dozens of edges need to be sliced for acceptable size in large tensor networks from big circuits like Sycamore\cite{google-nature-2019} , which makes multi-edge slicing a critical problem for complexity analysis. Considering that \emph{lifetimes} of different edges are independent, we can analyze the \emph{lifetime} of each sliced edge sequentially. Fig.~\ref{fig5} provides an example of the slicing of two edges with index $A$ and index $B$. From left to right, the contraction tree is divided into $5$ parts, lifetime of index $A$ goes through part $2$ and $3$, while $B$ goes through part $3$ and $4$. Index $A$ and $B$ double the time complexity of part $1, 4, 5$ and part $1, 2, 5$ respectively, and make the whole multiple of redundant calculation shown in Fig. \ref{fig5} as a piecewise production. Expanding to a more general circumstance, if a tensor contains $s$ sliced indices with $n$ in total, it will be divided into $2^s$ smaller tensors. For all $2^n$ subtasks, there will be $2^{n-s}$ times of redundant data. Then the time complexity will be multiplied by $2^{n-s}$ times. As a result, the density of \emph{lifetimes} plays a decisive role in memory and computation complexity in a region.

More sliced edges tend to lead to higher overhead. As a simple condition, when a slicing set is given, if we additionally pick more edges to slice, then the overhead will grow unless the \emph{lifetimes} of the added edges go across the whole contraction tree, which rarely happens. However, for an arbitrary slicing set with $n$ edges and $n+1$ edges, can we find a partial ordering relation between their overhead? We can prove the theorem below with a given target size. It provides interpretation for redundant slicing more profoundly and casts the theoretical cornerstone for our slicing strategy. It shows that a valid smaller slicing set indicates a lower slicing overhead in most conditions.
\begin{theorem}
    \label{theorem1}
    For a tensor network $G(V, E)$, a contraction tree $B(V_B, E_B)$ and an $n$-edge slicing set $S_1$, if an $(n-1)$-edge slicing set $S_2$ is found, and the intersection of $S_1$ and $S_2$ is not empty, then there must exist an $(n-1)$-edge slicing set $S_3$, whose overhead is less than $S_1$.
\end{theorem}
\begin{figure}[htbp]
\centerline{\includegraphics[width=0.52\textwidth,height=0.3\textwidth]{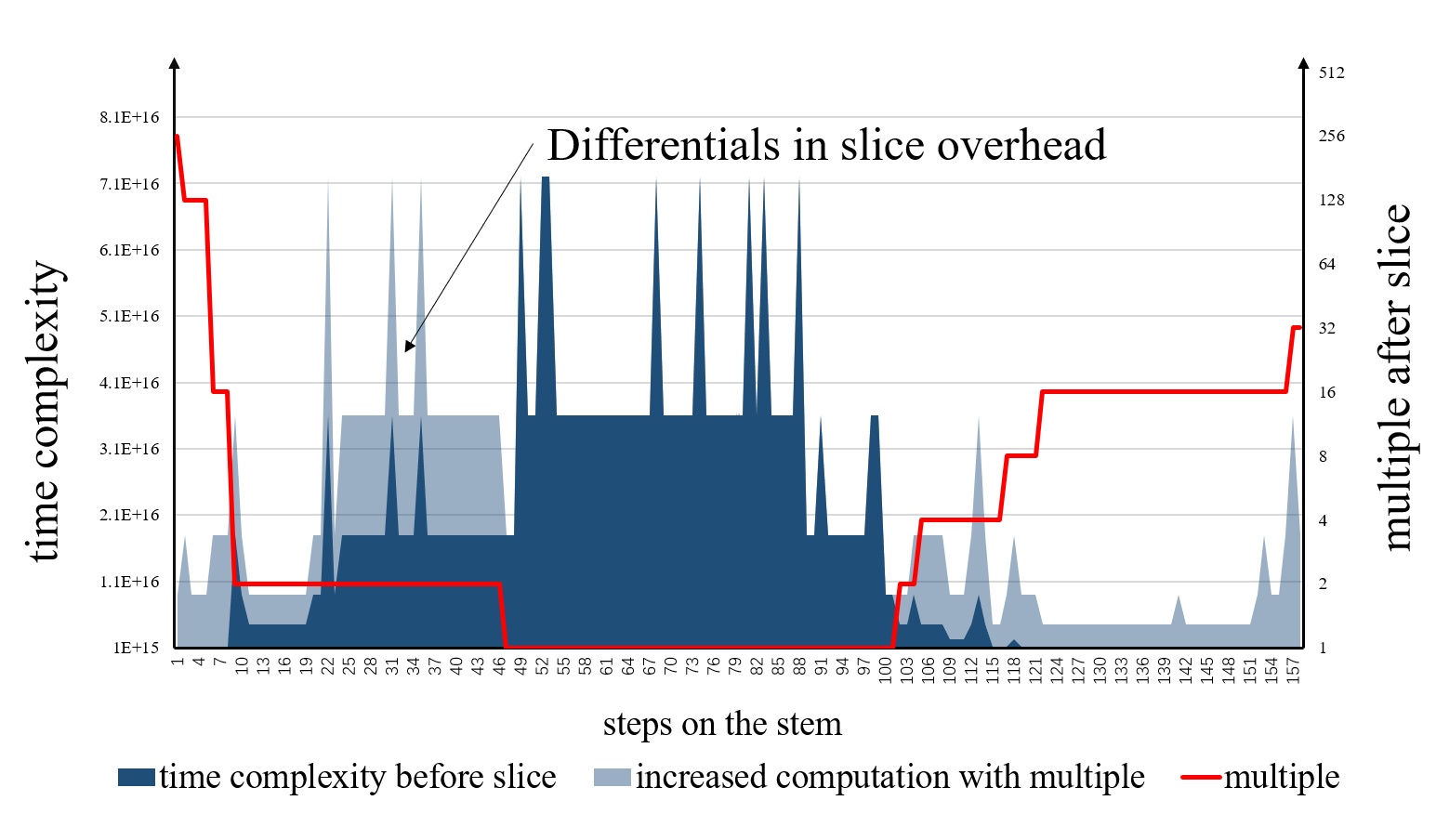}} 
\caption{\textbf{Time complexity and multiple by slicing on stem (Sycamore$(m=20)$ is used as an example).}}
\label{fig6}
\end{figure}


Fig.~\ref{fig6} compares the whole time complexity before and after slicing on a common contraction tree of Sycamore circuit\cite{google-nature-2019}. From the figure, we can conclude that the key to a low overhead is that the time complexity of the main computation-intensive part is kept, which means big tensors are contained in lifetimes of as many as sliced edges. Ideally, time complexity and the slicing caused by multiple are negatively correlated.

\subsection{Route of Slicing Optimization}
Due to the memory bound, we have to do slicing and stand with some overhead. However, on a multi-level storage system, under some circumstances, the computation overhead can be translated into data movement. 

For TNC, we will introduce two main methods to deal with the overhead, \emph{i.e.,} to reduce overhead by searching for better slicing sets or to replace overhead by data movement via stacking, in the following sections, respectively. Stacking is the inverse operation of slicing, and is feasible when the capacity of the low storage or distributed memory is enough for the whole memory requirement. For example, we can store the rank-53 tensor in the hard disk, get a rank-30 slice at a contraction step by IO, and then put it back after calculation. This process performs slicing by getting and stacking by putting naturally. If stacking immediately after the \emph{lifetime} of a sliced edge ends, the overhead of this edge is naturally eliminated, and the memory demand is resolved. Nevertheless, there will be vast costs of data exchange during data access (use low-level storage) or communication (for distributed memory). Searching for a better slicing set can avoid memory access and communication, but it introduces overhead from the redundant calculation. The choice depends on the specific value of bandwidth and overhead. In case of low bandwidth and low overhead, slicing optimization has better performance, while stacking is more suitable in case of high bandwidth and high overhead.

\begin{figure}[htbp]
\centerline{\includegraphics[width=0.5\textwidth,
height=0.323\textwidth]{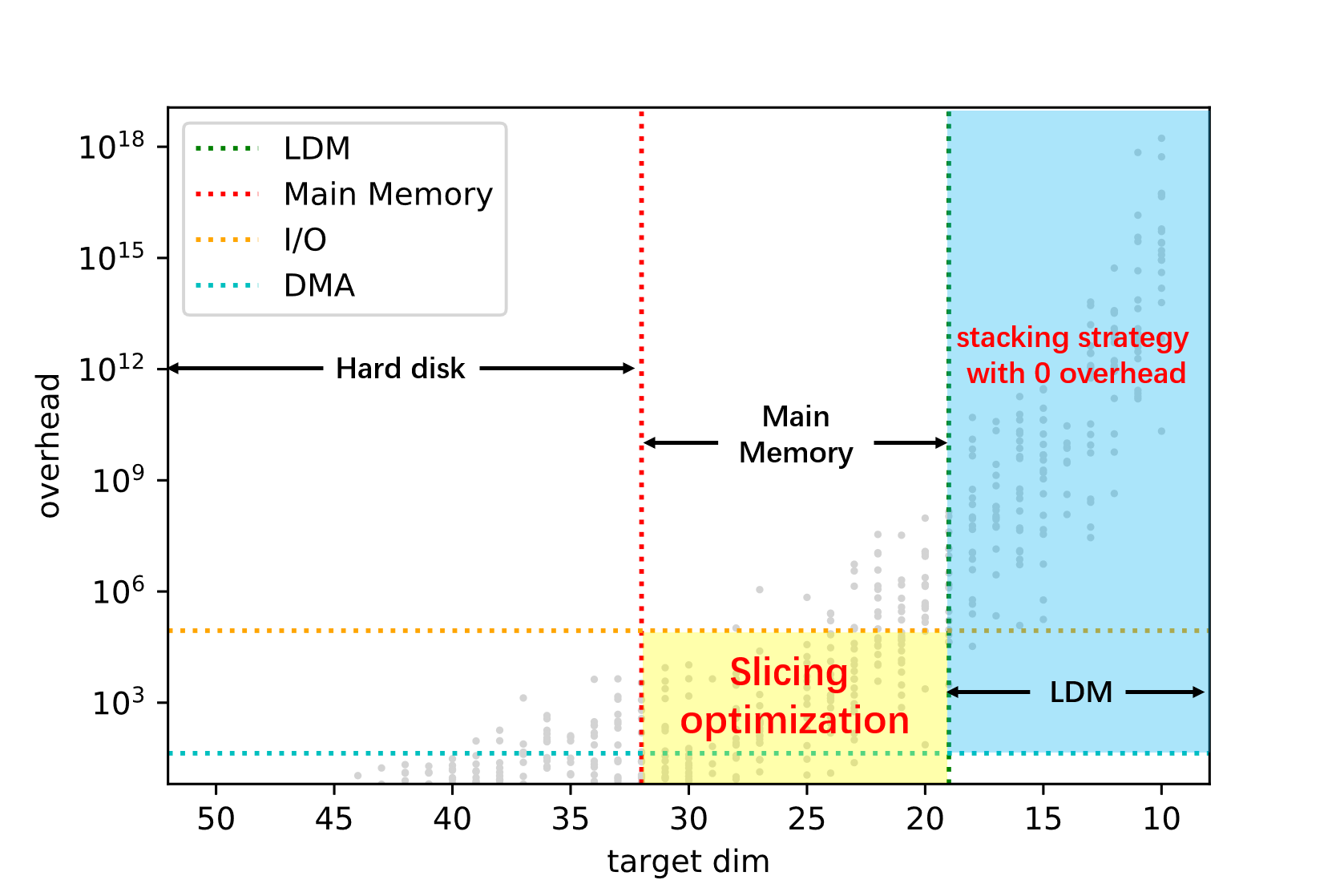}} 
\caption{\textbf{Overhead distribution for different storage level by cotengra. (Sycamore$(m=20)$ as an example with original memory cost dozens of PBs. For Sunway architecture, there is a 96GB main memory and a 256KB LDM for each CPE)}}
\label{strategy}
\end{figure}

Fig.~\ref{strategy} shows a typical overhead distribution for different target sizes. Data access costs are translated into equal overhead by arithmetic intensities (AI) of different levels. The line of equal overhead and the storage capacities divide the graph into several parts.  The commonest manually controllable storage levels include hard disk, main memory and LDM (for Sunway architecture). For memory access, $BandWidth_{IO} \ll BandWidth_{DMA} \ll BandWidth_{access LDM}$. Then, from hard disk to LDM, stacking becomes more potential, and in contrast, slicing optimization should be applied. 

These methods are not customized for a certain architecture. Actually, all we need is a multi-level storage system.

\section{Slicing Overhead Reduction\label{slicing strategy}}

\subsection{Overview of Slicing Optimization}
The Low bandwidth of IO restricts stacking, and the slicing overhead directly influences the parallel scalability. 
As shown in Theorem. \ref{theorem1}, optimization of slicing can be divided into two steps: searching for a smaller slicing set, and tuning the selected slicing set to achieve a lower overhead. The following two subsections discuss these two steps separately. 

Minimizing the overhead while fitting memory bound describes an optimization problem with constraints. For a contraction tree $B = (N_B, E_B)$, the target function can be described by the total time complexity after slicing.
\begin{equation}
C(B, S) = \sum_{V \in N_B}{2^{|s_V| + |S| - |S \cap s_V|}}
\end{equation}
where $S$ is the set of sliced edges, and $s_V$ is the set of edges which involve the contraction of $V$ in the TN.

The target size which does not exceed the memory capacity is denoted as $t$. For every $e \in E_B$, $|S \cap s_e| \geq |s_e| - t$. The target function and the formula together can determine the optimization problem. 

\subsection{Slicing Strategy}

Here we define \emph{lifetime} on the stem\cite{huang2020classical}, while the definition of stem is somehow different. In this work, stem is defined as the most computationally intensive path on the contraction tree. According to graph theory, on a path, each contraction step has data dependence with its next step. Since most branches have nothing to do with the memory constraints, they are pre-contracted, and only stems remain for optimization. After precondition, the stem becomes a new contraction tree, and we can detect \emph{lifetime} of every edge on it.

Assuming the optimal size of the slicing set is $s$, the stem is then separated into two continuous parts: $M = M_1 + M_2$, and let the optimal size of $M_1$ be $s_1$. For $M_1$, a number of candidate sets share the same size $s_1$. According to the discussion above, the more tensors \emph{lifetimes} of $s_1$ pass through $M_2$, the fewer edges shall be sliced in the updated $M_2$. 

Strictly speaking, longer \emph{lifetime} does not necessarily lead to lower slicing overhead. However, the containment relationship of \emph{lifetime} does. So is the effect of reducing memory. In particular, for the leaf nodes of the contraction tree, there is only one extension direction, then its length decides the containment relationship. So we start the search from the two ends of the stem, and choose one tensor as $M_1$ each time and the left part as $M_2$. Therefore, an edge with longer \emph{lifetime} shall be better. If we perform this process iteratively, a smaller slicing set is expected. We propose Algorithm. \ref{alg1} for details.


\begin{algorithm}
	\caption{Procedure slice finder}
	\label{alg1}
	\renewcommand{\algorithmicrequire}{\textbf{Input:}}
	\renewcommand{\algorithmicensure}{\textbf{Output:}}
	\begin{algorithmic}
		\REQUIRE Stem of contraction tree $M$, target dim $t$
		\STATE $N = len(M), S \leftarrow \emptyset, lf = dict()$
		\FORALL{edge indices in $M$}
		\STATE Calculate $lf[index]$
		\ENDFOR
		\REPEAT
		\STATE $sT = (M[0].dim < M[N-1].dim) ? M[0] : M[N-1]$
		\STATE Add $sT.dim - t$ indices with longest lifetime into $S$
		\FORALL{$T \in M$}
            \IF {$T.dim \leq t$}
            \STATE {$M.remove(T)$}
            \ENDIF
            \ENDFOR
		\STATE Update $lf$, $N$
		\UNTIL $N == 0$
		\ENSURE slicing set $S$
	\end{algorithmic}
\end{algorithm}

Though the above strategy does not always provide the slicing set with the lowest overhead, sometimes even worse than the one found by contengra, we have successfully satisfied all the prerequisites of the theorem above. The method above can find a slicing set as small as possible for a given contraction tree. Combined with the dynamic process for path searching, the overhead may drop lower. Then, the refiner will find a better slicing set to reduce the overhead.

\subsection{Slice Refiner based on Simulated Annealing}
Our slice finder can not guarantee to find an optimal slicing set, and it looks for one as small as possible. So, the main task of this section is to find the best one given a certain number of sliced edges. Considering that the size of the slicing set is fixed, the permutation from the former to the latter can be described as edge replacement. 

After slicing, there are some tensors whose dimensions are exactly equal to the target dimension, and they are named "critical tensors". If the lifetime of a sliced edge contains none of the critical tensors, this edge does not contribute to memory reduction, and can be removed from the slicing set. Moreover, from another angle of view, to replace a sliced edge with index $a$, a direct way is to find another unsliced index $b$ whose lifetime contains all critical tensors in the lifetime of $a$. 

Simply replacing indices greedily sometimes leads to a local minimum. For instance, the optimal set contains index $a$ and index $b$ while the one we found contains $c$ and $d$, but replacing $c$ from $a$ may increase the overhead, or fail to match the memory bound. Exhausted searching or dynamic programming will lead to the optimal solution, but suffers from a combination explosion of search space. Then, stochastic algorithm becomes a choice to balance the time and quality of searching. Simulated annealing endures a temporary complexity increase to avoid local minimum, and provides a practical speed, as the Algorithm. \ref{alg2} shows.

\begin{algorithm}
	\caption{Procedure slice refiner}
	\label{alg2}
	\renewcommand{\algorithmicrequire}{\textbf{Input:}}
	\renewcommand{\algorithmicensure}{\textbf{Output:}}
	\begin{algorithmic}
		\REQUIRE Target dim $t$, slicing set $S$, contraction tree $M$, final temperature $tf$, parameter $\alpha$
        \STATE Initialize temperature $T$
		\FORALL{edge indices in $S$}
		\STATE Calculate $lf[index]$
		\ENDFOR
		\REPEAT
        \STATE Randomly choose $index$ from $S$
        \STATE $c\_tensors = find\_critical\_tensors(lf[index], t)$
		\STATE $candidate = find\_candidate\_indices(c\_tensors)$
        \FORALL{$can \in candidate$}
            \STATE Replace $index$ by $can$
            \STATE Calculate time complexity $C_{new}, C_{ori}$
            \IF{$C_{new} < C_{ori}$}
                \STATE Update $S$
            \ELSE
                \STATE Update $S$ with prob. $p = exp((C_{ori} - C_{new})/C_{ori}/T)$
            \ENDIF
        \ENDFOR
		\UNTIL $T < tf$
		\ENSURE slicing set $S$
	\end{algorithmic}
\end{algorithm}

\section{Fused Design by Secondary Slicing}
\subsection{Difficulty for Thread Level Optimization}
For thread level, if organizing the whole path on the CPEs, considering the 256-KB LDM, there will be a huge overhead. As a result, previous works optimize on thread-level step by step\cite{li2021sw_qsim}\cite{liu2021closing} to avoid it, turning to memory access instead of vast redundant calculation. However, frequent memory access also limits the FLOPs efficiency a lot.

Contraction between tensors is implemented by matrix multiplication\cite{li2021sw_qsim}. Based on this strategy, more than $70\%$ of the peak performance can be obtained for square-like matrices. However, for narrow matrix multiplication, in particular, when two of the $m, n, k$ are less than 16, which dominated many simulation tasks\cite{google-nature-2019}, we can easily deduce $\Theta(MNK) \approx \Theta(MN+NK+MK)$, so that GEMM will change from a computational intensity problem to a bandwidth-constrained problem. Another constraint comes from the 256-KB LDM on each CPE, which can only hold a rank-13 tensor for multiplication. Then, calculating a contraction by a rank-31 tensor at least needs $2^{12}$ times low-efficiency GEMM for a CG.

To take full advantage of the 512-bits single instruction multi-data (SIMD), a $4\times 4$ complex GEMM kernel is applied. When $k, n$ are both less than $4$ in one thread, the demand of data length will cause a conflict between the high-performance kernel and the 2D data distribution. In contrast, 1D distribution avoids most padding, and slices the GEMM process into simply paralleled subtasks. Most importantly, the implicit slicing embedded in data distribution, as a method to overcome the memory limitation of LDM, has a similar problem with slicing on the process level.

We propose a more efficient slice-stack strategy guided by \emph{lifetime}, which can heavily reduce the cost of data movement. Specific for thread level, it is named as \textbf{secondary slicing} (first slicing happens on the process level). Secondary slicing between LDM and the main memory will provide a fused operator to organize parts of the contraction path on CPEs, and greatly save the memory access cost.
\subsection{Slice-stack design}
\begin{figure*}[htbp]
\centerline{\includegraphics[width=1.0\textwidth,height=0.6\textwidth]{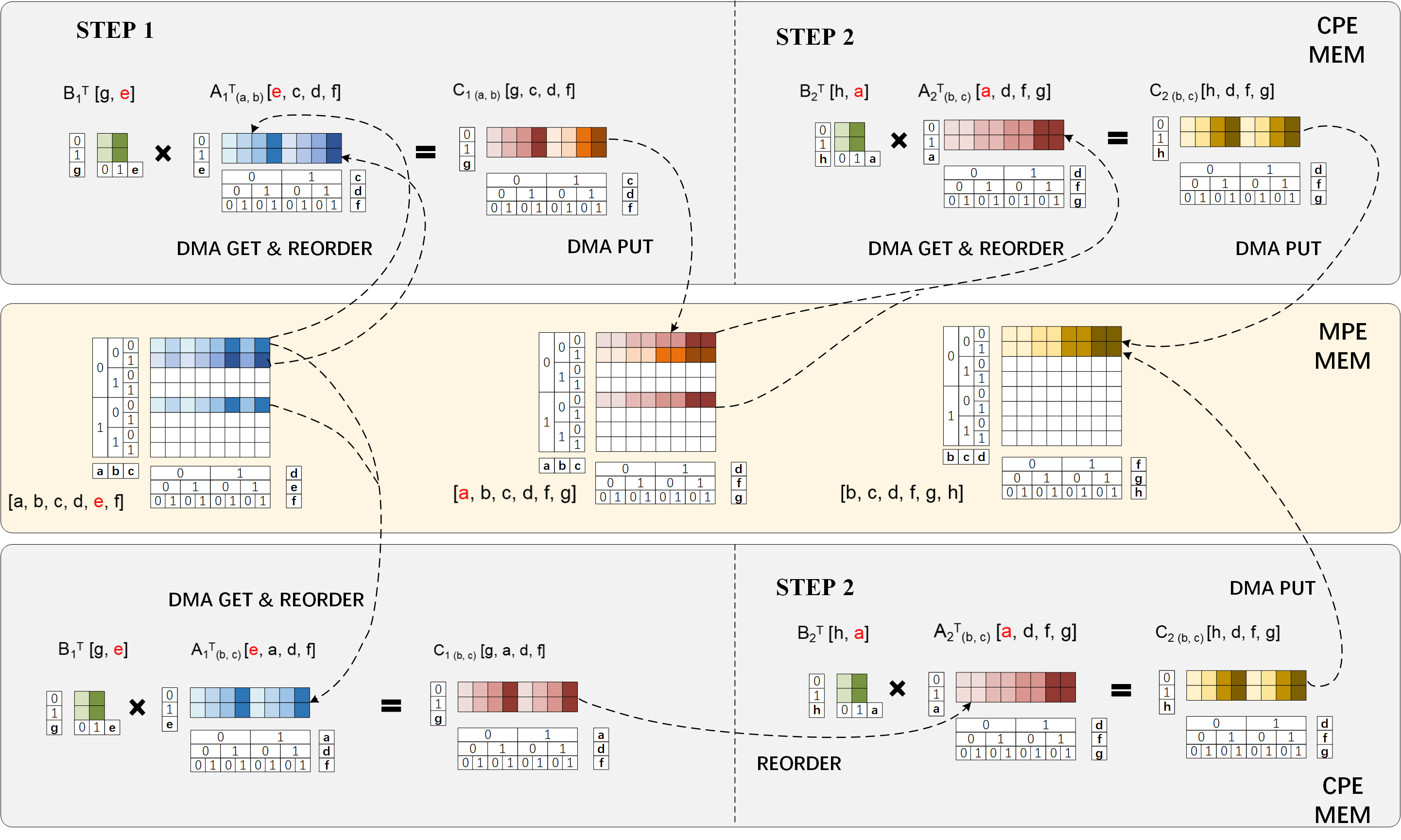}} 
\caption{\textbf{A fused design for thread-level optimization. The top part shows the step-by-step strategy in previous works, and the bottom is the fused design. Our strategy organizes a sub-path on the LDM, and reduces $n-1$ times of DMA-get and DMA-put for a length-$n$ sub-path. The superscript $T$ denotes permutation, and the subscript of tensors denotes the sliced dimensions.}}
\label{fig8}
\end{figure*}
If we do slicing on the whole contraction path, a subtask can be performed on a process without intermediate communication and IO while standing with computation overhead. In contrast, if we accept the communication or memory access, the computation overhead can be avoided. 

Such a strategy does not work on the process level due to the low bandwidth of IO and low computation overhead, but is available between LDM and the main memory. Since LDM on the new Sunway supercomputer is only feasible for a rank-13 tensor, if the contraction is performed step by step, we should frequently do DMA-get and DMA-put between every two steps. Considering the arithmetic intensity above, memory access will be dominant. To reduce the frequent data exchange between LDMs and the main memory, performing $n$ steps of the contraction path successively in one computation kernel helps. At the beginning of those steps, we do DMA-get once to get the bigger tensor, and after n steps of calculations, the result will be written back by DMA-put. Then, $n-1$ times of DMA-get and DMA-put are reduced.

As the top part of Fig.~\ref{fig8} shows, CPEs can obtain a lower rank tensor from the one in the main memory at each time, throwing out some dimensions. In a particular contraction, commonly, there are parts of dimensions absorbed, and the others held. According to this observation, at the beginning of the computation, if we choose all dimensions which will be absorbed in the following $n$ steps to form a tensor, and get it into LDMs by DMA with stride, then each CPE can do the calculation independently, like the bottom part of Fig.~\ref{fig8} shows. As for the memory bound, we tune the parameter $n$ to determine the rank of the formed tensor.

In this strategy, the chosen indices remain, and the other indices are sliced. Different CPEs do different subtasks generated by slicing, which are embarrassingly parallel. The main question is how to choose sliced indices and find a proper $n$. The fundamental prerequisite for the slicing set is that the sliced indices will not be contracted in $n$ steps, and fortunately, such a condition is just the definition of \emph{lifetime}. So with a particular start position on the path, we slice the indices which have the longest \emph{lifetime} and traverse the stem until the \emph{lifetime} of any one of the sliced indices ends. If the size of the slicing set added by $13$ (the capacity of LDM) is less than each tensor on the sub-path, the fused calculation is finished. With this strategy, there's no slicing overhead, and the last DMA-put plays the role of stacking with no extra communication.
\begin{figure}[htbp]
\centerline{\includegraphics[width=0.5\textwidth,height=0.14\textwidth]{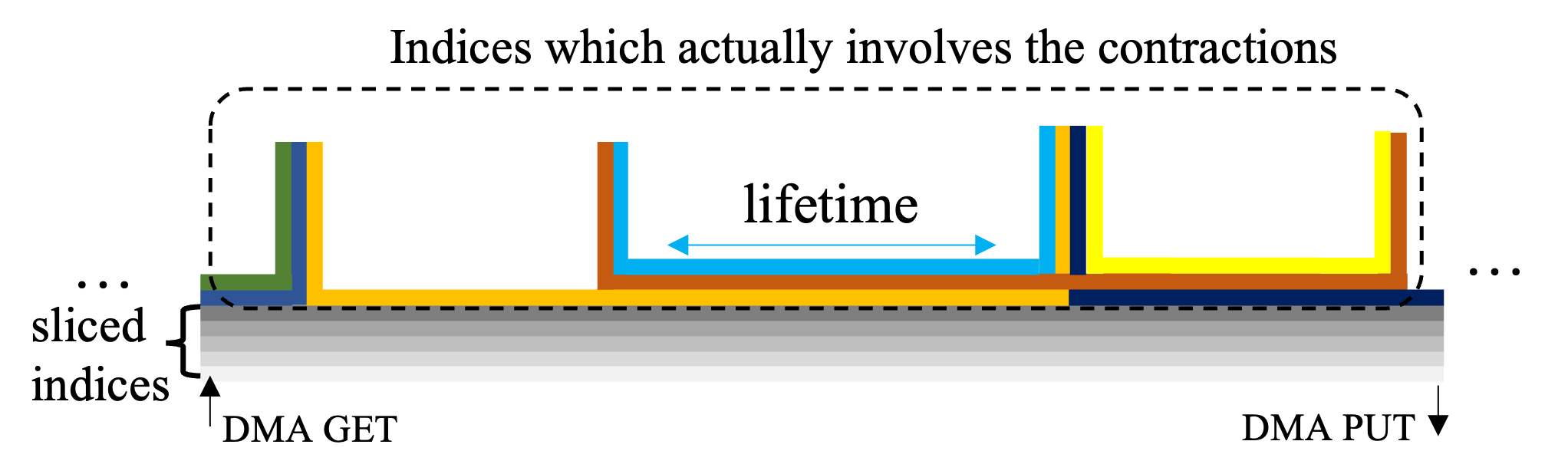}} 
\caption{\textbf{An example of secondary slicing. There are 5 edges whose \emph{lifetime} go across the whole. After slicing these edges, the maximum memory demand is reduced from $2^8$ to $2^3$, and $2^5$ subtasks are generated for thread-level parallel. Only the computation inside the dotted box will be performed in each core, and memory access happens at the beginning and the end of the computation.}}
\label{secondary slicing}
\end{figure}

\subsection{Deep Optimization}
\subsubsection{Permutation Map Reduction by Recursion Formula}

In situ computing map and the pre-calculated map are two primary methods for the tensor permutation. The in situ computing map needs an $O(NlogN)$ time complexity for a size-$N$ tensor with rank $O(logN)$, and an $O(1)$ space complexity. In contrast, the latter provides $O(NlogN)$ time complexity for the first computation and $O(N)$ for the next ones, and space complexity of $O(N)$. In our fused algorithm, permutations before every step of contractions become one of the hot spots. Even though using short type, $n$ pre-calculated maps are too big to store. However, for the strategy based on in situ map, the tensors' ranks are about 10, which leads to more than 10 times the cost.

A feasible solution is to combine the advantage of both strategies. For a specific contraction, $A^TB^T = C$, in which $T$ denotes permutation, a typical indices order of a rank-9 $A$ after permutation is like $0, 1, 2, 4, 5, 7, 8, 3, 6$, the indices need to be absorbed will be organized at the end, \emph{i.e.}, $3, 6$. Contrary to $A$, such indices are placed at the beginning, like $3, 8, 0, 1, 2, 4, 5, 6, 7$. For $A$, obviously, the first 3 dimensions will not participate in the permutation, so only a $1/8$ map is enough. And for $B$, the last 4 dimensions are the same during the permutation, so the adjacency of the 4 elements shaped by $4,5,6,7$ will be held, and the size of the map is reduced to $1/16$. With an offset, map of the corresponding can be calculated by $map[i+k] = map[i]+k*offset$ when $k < stride$ in $O(1)$, the same as pre-calculated map. The space complexity will be reduced to $O(N/2^m)$, where m is the number of continuous indices at the end or beginning.

\subsubsection{Eliminate Discrete Memory Access by Communication}

Though the fused strategy heavily reduces DMA, the efficiency of DMA becomes an obstacle. Since we have sliced some edges from the original tensors, the sub-tensors we need are often discretely distributed in the main memory. If we want to get a sub-tensor without the last index (assume that it has been sliced), there will be a space interval between every two elements we need. As a result, the granularity of DMA will be extremely small, which harms the bandwidth. In practice, the sliced edges are scattered in different positions. Under this circumstance, the bandwidth of DMA can only achieve less than $0.1\%$ of the peak performance, and makes negative optimization. 

To improve the bandwidth, the cooperation of 64 CPEs becomes essential. Under our parallel framework, the last 6 sliced indices are organized between 64 CPEs. Data exchange is much faster between CPEs by Remote memory access (RMA), whose peak performance can achieve 800GB/s for a whole CG. Then, we treat the sub-tensors of a 64 CPEs as a whole, and let every CPE access the memory continuously. This strategy guarantees a basic granularity of 512B for DMA, which provides more than $50\%$ of the peak performance. After memory access, RMA is applied to rearrange the data between CPEs. An additional permutation is applied to improve the granularity of RMA and ensure efficiency. For other architecture, we can implement this strategy by simply partitioning the communication groups.

\section{Result}
Related efforts (\emph{e.g.}, Cotengra\cite{gray2021hyper}, Alibaba\cite{huang2020classical} and \cite{liu2021closing}) have all simulated Sycamore RQC\cite{google-nature-2019}. To make an appropriate comparison, unless otherwise specified, the contraction trees used in this work come from the tensor network of Sycamore as well.

\subsection{Slicing Overhead and Scaling Result}
\begin{figure}[htbp]
\centerline{\includegraphics[width=0.5\textwidth,
height=0.33\textwidth]{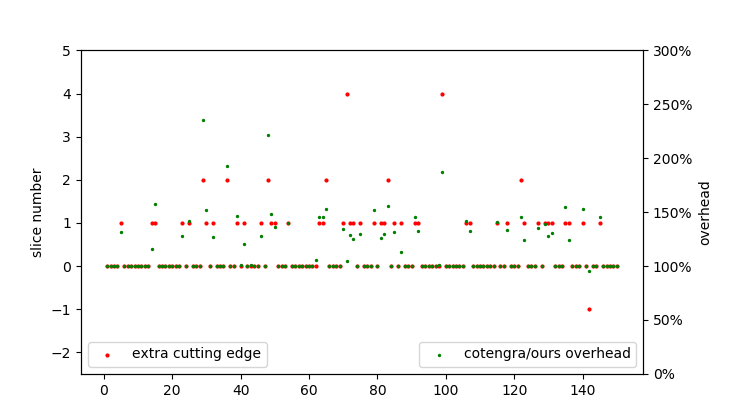}} 
\caption{\textbf{Slicing size and overhead compared with cotengra. The red points shows the number of extra slicing edges by cotengra compared with ours; and the green points shows the ratio of overhead. When the difference is $0$ and the ratio is $100\%$, two method perform equally.}}
\label{fig9}
\end{figure}
According to the slice strategy in section \ref{slicing strategy}, given a contraction path, we can find a slicing set with lower overhead in expectation. Fig.~\ref{fig9} shows the comparison between our work and cotengra. We find 400 contraction paths by contengra, and apply our method and cotengra respectively to search for slicing sets. Compared with cotengra, the slicing sets we found are potentially smaller, and lead to lower overhead. Specific to each path, our strategy performs better on more than $98\%$ of cases. As for several exception cases, an observed phenomenon is that the computationally intensive regions of the paths are dispersed. And also, since SA is a stochastic method, it can not guarantee to find an optimal solution.

The best overhead result we found is less to 1.05, and we comprehensively choose a path with low complexity to do the thread level optimization.

\begin{figure}[htbp]
\centerline{\includegraphics[width=0.5\textwidth,
height=0.28\textwidth]{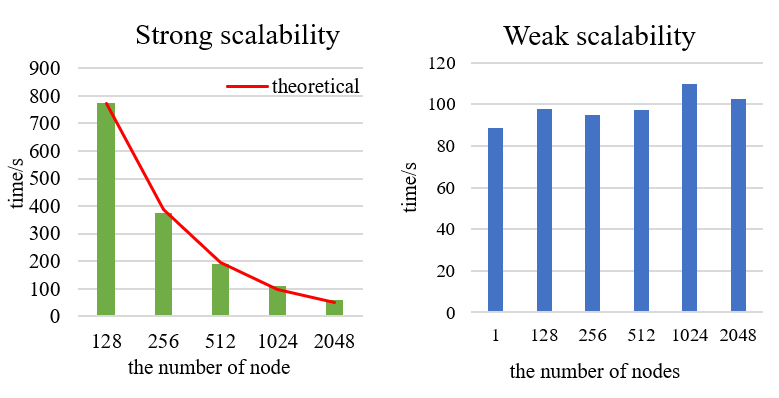}} 
\caption{\textbf{(Strong scaling results (65536 subtasks in total) and weak scaling results (16 subtasks on each node).}}
\label{fig10}
\end{figure}

Due to slicing, processes are highly independent with each other, and will only do \emph{allReduce} once at the end of program. The strong scaling results and weak scaling results are showed in Fig.~\ref{fig10}. 

\subsection{Computing Efficiency}

\begin{figure}[htbp]
\centerline{\includegraphics[width=0.5\textwidth,
height=0.34\textwidth]{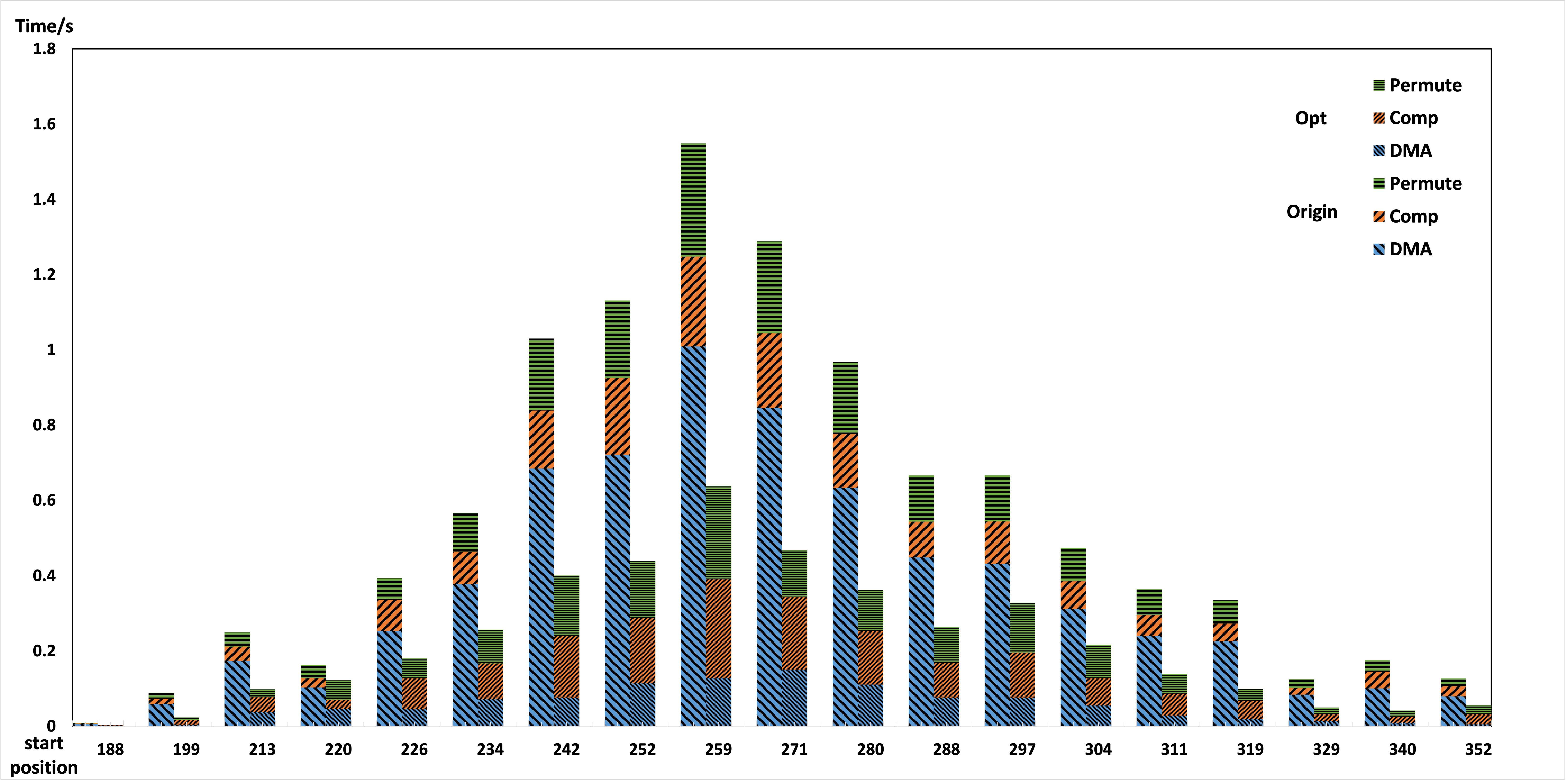}} 
\caption{\textbf{Optimization by secondary slicing at the thread level. Optimization is done on a single node with \textbf{$390$} cores, and the performance is tested for tasks with different size on a contraction path. Use the step-by-step strategy for comparison.}}
\label{fig11}
\end{figure}

Fig. ~\ref{fig11} shows that our work is able to significantly improve the computing efficiency. Using 1024 nodes, a perfect sample or 1M correlated samples can be generated in 10098.5s. Considering about the scaling result, we project that we can reduce the whole time cost using 107520 nodes (41,932,800 cores) to 96.1s. The sustainable single-precision performance is projected as 308.6Pflops.

\begin{figure}[htbp]
\centerline{\includegraphics[width=0.5\textwidth,
height=0.34\textwidth]{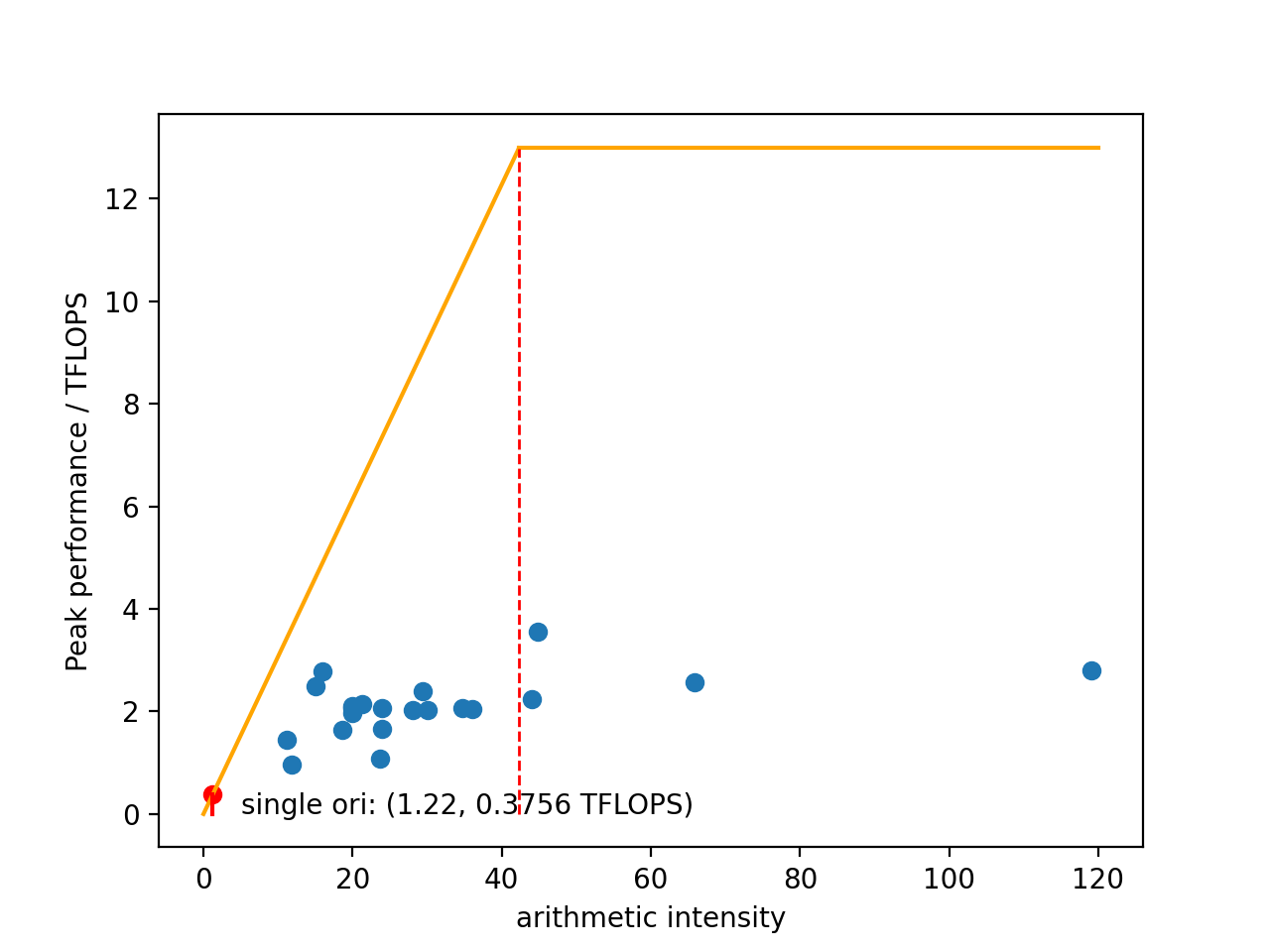}} 
\caption{\textbf{Roofline Model of our work. For different cases, the arithmetic intensities are different from $10x$ to $40x$. In some cases, the problems turn into computation-bound.}}
\label{fig12}
\end{figure}

According to Fig. ~\ref{fig11}, the time of memory access is largely reduced by secondary slicing, and the time of permutation and GEMM keep similar. This result verifies our prediction that, secondary slicing can reduce memory access with the price of some python based pre-conditioning, which needs only 1 core and ignorable time. Another conclusion is that, after secondary slicing, the computation kernel is transformed from a memory intensive one to a computation intensive one in some cases. According to Roofline model\cite{osti_1407078} and the arithmetic intensity of Sunway architecture, the number of floating point operations should exceed 42.3 times of number of bytes of memory access to achieve a peak performance. The average fused steps of our strategy is about 10, and there is $MNK / (MN + MK + NK) \approx K$ for a small $K$ with a average of 4, then in some cases the arithmetic intensity of TNC can break through 42.3. Fig. ~\ref{fig12} shows the Roofline model for a particular case on thread level. Due to permutation, which consists of LDM access, there is still a gap between our performance and the peak performance. However, compared with the original arithmetic intensity of $2.6$ (mixed precision) and $1.22$ (single precision), the optimization limit is largely improved.

\section{Implication}
In this paper, we presented novel \emph{lifetime}-based methods to reduce the slicing overhead and improve the computing efficiency for parallel optimization at both the process level and thread level. 

Due to the introduction of \emph{lifetime} definition, A series of lemmas and theorems based on \emph{lifetime} intensify our awareness of a tensor network and the contraction trees, then an interpretable method to deal with slicing overhead, and an in-place slicing strategy to find the smallest slicing set, so that we can reduce the slicing overhead. Our in-place strategy can work independently, and can also optimize the slicing sets found by dynamic methods. As a result, it becomes a general post-process when searching for slicing sets.

Furthermore, based on \emph{lifetime}, we adopted stacking to avoid heavy overhead from the redundant calculation, and generalize it for arbitrary multi-level storage systems. Slicing optimization and stacking make up the two main strategies for TNC. Then, we design a discriminant based on arithmetic intensity to choose the strategy on different hierarchies of memory. For Sunway architecture, we applied it at the thread level. At the necessary price of more affordable memory access by DMA and faster communication among the CGs by RMA. The most inspiring result is that, in some cases, the computation kernels are changed from memory-intensive into computation-intensive, which reshapes our understanding of the problem.

As a widely applied approach, tensor networks can be found in many research fields like statistical physics\cite{evenbly2015tensor}, data science\cite{cichocki2017tensor}, sociology\cite{porter2009communities}, and so on. A series of problems in physics can be reduced to TNC. For large TNC, the extreme time and space complexity become the major difficulty. Translating the huge memory demand into a data stream that can be solved as a pipeline, slicing plays a significant role during these TNC processes. In addition to TNC, if we need to deal with a series of large matrix multiplications which have data dependence, slicing can still work for task distribution.

This work proves that \emph{lifetime} is essentially promising for the tensor networks of quantum circuits. Moreover, with the simple definition, \emph{lifetime} can be easily generalized to tensor works with different features and helps analyze both the time- and space-complexities of the contraction. It is foreseeable that \emph{lifetime} can play important roles in more fields.


\section*{Acknowledgment}

We would like to thank Zhen Wang, Zhaoqi Sun, Zegang Li and Yuxuan Li for advice and discussions. 

This work is partially supported by National Key R\&D Program of China (2020YFB0204804, 2020YFB0204800), National Natural Science Foundation of China (Grant No. T2125006, U1839206, Project No. 62102114), Jiangsu Innovation Capacity Building Program (Project No. \\ BM2022028) and the Key Research Project of Zhejiang Lab (No. 2021PB0AC01). The corresponding authors are Yong Liu, Xin Liu, Lin Gan, Dexun Chen and Guangwen Yang.

\bibliographystyle{ieeetr}
\bibliography{ref} 
\end{document}


\title{Supplementary Information of Lifetime-based Optimization for Simulating Quantum Circuits on a New Sunway Supercomputer}
\date{\today}

\maketitle

\section{Properties of lifetime}

With the definition of \emph{lifetime}, some more definitions and lemma should be introduced for derivation.

\begin{definition}
Given a contraction tree $B = (N_B, E_B)$, the concept correlated contractions of index $k$ refers to a set of nodes(contractions) $\{N_{i_1}, N_{i_2}, \dots, N_{i_m}\} \subset N_B$, if $k \in s_{N_{i_j}}$, for all $1\leq j\leq m$.
\end{definition}
\vspace{0.5em}

Based on above definitions, \emph{lifetime} has some good properties, and the most important one is linearity. 
\vspace{0.5em}
\begin{lemma}
\label{lemma1}
\textbf{Conservation} Given a contraction tree $B = (N_B, E_B)$, for any index in $s_{E_B}$, if it involves the corresponding contraction, it will only be in the two incidence edges of a node. Otherwise, it will not exist at all in $s_{node}$. 
\end{lemma}
\vspace{0.5em}

Conservation describes a common condition that an index involves a contraction. If contracted, it will only exist in the two contracted tensors; otherwise, it will be kept from one of the contracted tensors to the generated tensor. An interesting corollary from this theorem is that, contractions do not create or diminish index. Instead, they connect two tensors that share common index. Another inference is that, a correlated contraction of an index contains two tensors which exist in the indices' \emph{lifetime}. Furthermore, for every index, all contractions of the tree involved by tensors in its \emph{lifetime} form a correlated contraction set. Therefore, \emph{lifetime} and correlated contractions demonstrate a linear structure:

\vspace{0.5em}
\begin{theorem}
\label{thm1}
\textbf{Linearity} Given a contraction tree $B$ and an index $k$, there exists a path on the tree from one leaf node to another (closed leg) or the root (open leg). The set of edges the path goes through contains the \emph{lifetime} of $k$, and the set of nodes on the path is equivalent to the correlated contraction set of $k$.
\end{theorem}

\textbf{Proof:} In the TNC process, the indices will not be created or destroyed without foundation. All the indices come from original TN. As a result, for a closed leg, there must be two leaf nodes which contains it, otherwise it will not be contracted and be reserved to the root. Then we can connect the two leaf nodes, or the leaf node and the root, which forms a path of the tree. According to Lemma. \ref{lemma1}, when we go through the path from each leaf node to the node where index $k$ is contracted, $k$ exists in one of the input tensors and the output tensor before contracted. When $k$ is contracted, $k$ exists in the two input tensors. When other contractions are performed, $k$ has not been introduced, or has been already contracted. As a result, the whole \emph{lifetime} of $k$ is contained by the path.

\section{Proof of theorem 1 in the main text}
We do the proof by mathematical induction.
Given a slicing set $S_1$ of size $n$ and another set $S_2$ of size $n-1$, the intersection of $S_1$ and $S_2$ is $S'$. 
For $n = 2$, considering about the discussion in section 3 of the main text, if there are two sliced indices, the overhead caused by each index will be multiplied by parts. When $S' \neq \emptyset$, \emph{i.e.} $S_2 = {a}, S_1 = {a,b}$, additional overhead will occur at the regions where \emph{lifetime} of $b$ has not covered. As a result, the overhead of $S_1$ is larger than $S_2$. Moreover, we can achieve the conclusion even with an empty $S'$. Since \emph{lifetime} is linear, the tensors exceed the target size are gathered as a line on the \emph{lifetime} of the only edge $c$ in $S_2$. Assume that the \emph{lifetime} of $c$ is extremely short that it can only cover the line shaped by those tensors. As for edge $a, b \in S_1$, each of $a, b$ can only cover part of the tensors by one direction, and can reach the maximum length when their \emph{lifetime} cover all tensors but one. Otherwise, another sliced edge will be redundant and provides more overhead. Then according to Fig 5 in the paper, the contraction path is divided into 5 parts, with multiple of $S_1$: \{2, 2, 1, 2, 2\} and $S_2$:\{2, 1, 1, 1, 2\}, which means $S_2$ performs better than $S_1$.

For $n \geq 2$, assume that the conclusion works when $n = k$, and when $n = k + 1$, there is:

Considering that $S'$ is not empty, we simply choose one edge $d$ from $S'$ to slice, and for the sliced tensor network and the same contraction path, we can apply the assumption of $n = k$. Since $S_1 - \{d\}$ is a size-$k$ set, and $S_2 - \{d\}$ is a size-$k-1$ set, there must be a $S_3$ who provides overhead lower than $S_1 - \{d\}$. Since $S_3 + \{d\}$ is a size-$k$ set, there will be $S_3 + \{d\}$ better than $S_1$. Then the assumption still work for $n = k + 1$.

\section{Details of the Slicing Strategy}
We have described the slicing strategy in the main text. Some details about the processes to find the critical tensors and the candidate indices are described below, as Algorithm. \ref{alg3} and Algorithm. \ref{alg4} shows.
\newpage
\begin{algorithm}
	\caption{Procedure find\_critical\_tensors}
	\label{alg3}
	\renewcommand{\algorithmicrequire}{\textbf{Input:}}
	\renewcommand{\algorithmicensure}{\textbf{Output:}}
	\begin{algorithmic}
		\REQUIRE Target dim $t$, lifetime of the chosen index $lf$
        \STATE Initialize $c\_tensors = \emptyset$
		\FORALL{tensor $T \in lf$}
            \IF {$T.dim == t$}
            \STATE $c\_tensors.append(T)$
            \ENDIF
		\ENDFOR
		\ENSURE set of critical tensors $c\_tensors$
	\end{algorithmic}
\end{algorithm}

\begin{algorithm}
	\caption{Procedure find\_candidate\_indices}
	\label{alg4}
	\renewcommand{\algorithmicrequire}{\textbf{Input:}}
	\renewcommand{\algorithmicensure}{\textbf{Output:}}
	\begin{algorithmic}
		\REQUIRE Set of critical tensors $c\_tensors$
        \STATE Initialize $candidate$ as the set of indices of $c\_tensors[0]$
		\FORALL{tensor $T \in c\_tensors$}
            \STATE $sT =$ the indices set of $T$
            \STATE $candidate = candidate \cap sT$
		\ENDFOR
		\ENSURE set of candidate indices $candidate$
	\end{algorithmic}
\end{algorithm}